\documentclass[conference]{IEEEtran}
\IEEEoverridecommandlockouts
\usepackage{cite}
\usepackage{amsmath,amssymb,amsfonts}
\usepackage{graphicx}
\usepackage{textcomp}
\usepackage{xcolor}
\usepackage{xspace}
\usepackage{hyperref}
\usepackage{tabu}
\usepackage{algorithm}
\usepackage[noend]{algpseudocode}
\usepackage{array}
\usepackage[T1]{fontenc}
\usepackage{enumitem}
\usepackage{multirow}

\usepackage{adjustbox}

\usepackage[noend]{algpseudocode}

\newcolumntype{P}[1]{>{\centering\arraybackslash}p{#1}}
\newcolumntype{M}[1]{>{\centering\arraybackslash}m{#1}}

\usepackage{pifont}
\usepackage{xcolor}
\newcommand{\cmark}{\textcolor{green!80!black}{\ding{51}}}
\newcommand{\xmark}{\textcolor{red}{\ding{55}}}

\newtheorem{definition}{Definition}

\def\BibTeX{{\rm B\kern-.05em{\sc i\kern-.025em b}\kern-.08em
    T\kern-.1667em\lower.7ex\hbox{E}\kern-.125emX}}

\usepackage{tikz,xcolor,hyperref}

\definecolor{lime}{HTML}{A6CE39}
\DeclareRobustCommand{\orcidicon}{%
	\begin{tikzpicture}
	\draw[lime, fill=lime] (0,0) 
	circle [radius=0.16] 
	node[white] {{\fontfamily{qag}\selectfont \tiny ID}};
	\draw[white, fill=white] (-0.0625,0.095) 
	circle [radius=0.007];
	\end{tikzpicture}
	\hspace{-2mm}
}

\foreach \x in {A, ..., Z}{%
	\expandafter\xdef\csname orcid\x\endcsname{\noexpand\href{https://orcid.org/\csname orcidauthor\x\endcsname}{\noexpand\orcidicon}}
}

\begin{document}

\title{Identifying Threats, Cybercrime and Digital Forensic Opportunities in Smart City Infrastructure via Threat Modeling\\
}

\author{\IEEEauthorblockN{Yee Ching Tok\orcidA{}}
\IEEEauthorblockA{\textit{Singapore Univ. of Tech. and Design} \\
Singapore \\
yeeching\_tok@sutd.edu.sg}
\and
\IEEEauthorblockN{Sudipta Chattopadhyay\orcidB{}}
\IEEEauthorblockA{\textit{Singapore Univ. of Tech. and Design} \\
Singapore \\
sudipta\_chattopadhyay@sutd.edu.sg}
}

\maketitle
\thispagestyle{plain}
\pagestyle{plain}

\begin{abstract}
Technological advances have enabled multiple countries to consider implementing Smart City Infrastructure to provide in-depth insights into different data points and enhance the lives of citizens. Unfortunately, these new technological implementations also entice adversaries and cybercriminals to execute cyber-attacks and commit criminal acts on these modern infrastructures. Given the borderless nature of cyber attacks, varying levels of understanding of smart city infrastructure and ongoing investigation workloads, law enforcement agencies and investigators would be hard-pressed to respond to these kinds of cybercrime. Without an investigative capability by investigators, these smart infrastructures could become new targets favored by cybercriminals.

To address the challenges faced by investigators, we propose a common definition of smart city infrastructure. Based on the definition, we utilize the STRIDE threat modeling methodology and the Microsoft Threat Modeling Tool to identify threats present in the infrastructure and create a threat model which can be further customized or extended by interested parties. Next, we map offences, possible evidence sources and types of threats identified to help investigators understand what crimes could have been committed and what evidence would be required in their investigation work. Finally, noting that Smart City Infrastructure investigations would be a global multi-faceted challenge, we discuss technical and legal opportunities in digital forensics on Smart City Infrastructure.
\end{abstract}


\section{Introduction} \label{Introduction}

\begin{figure}[ht]
\includegraphics[width=\columnwidth]{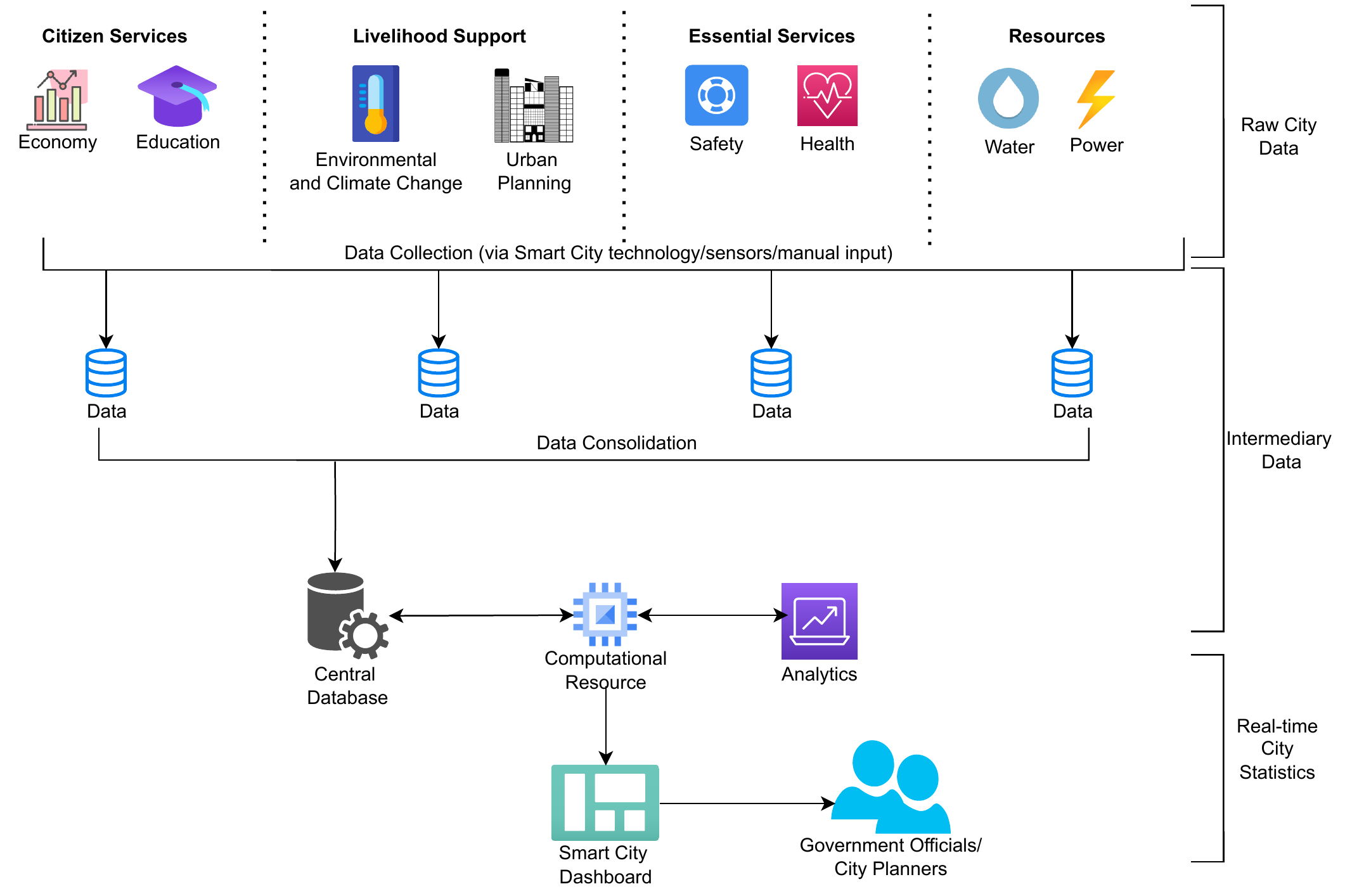}
\centering
\vspace{-15pt}\caption{Overview of Smart City Infrastructure (SCI) Data Collection and Processing}\vspace{-18pt}
\label{fig:1_SCI_Intro}
\end{figure}

In this increasingly interconnected world, humans generate a lot of data in their daily lives through the usage of computing devices and the multitude of increasingly accessible technologies that enable a better quality of life. The technologies include using smart home devices (popularly termed as Internet-of-Things (IoT)) and interaction with novel technical implementations by governments to improve citizens' quality of life. These novel technical implementations range from smart water and electricity meters, smart vehicles, autonomous vehicles and building automation systems~\cite{BaigZubair2017}. These governments' final goal was to presumably expand the systems into Smart City Infrastructure (SCI) for a better overview of their citizens, environment, safety and resources.~\autoref{fig:1_SCI_Intro} illustrates how governments and city planners could gather various data points to obtain a holistic overview of the country, allowing them to provide timely governance intervention where needed.

Unfortunately, governments and city planners are not the only ones embracing the adoption of technologies related to SCI. These new systems offer attractive opportunities to state-sponsored adversaries and cybercriminals. If SCI data could be stolen or intercepted, adversaries and cybercriminals could gain access to a plethora of data that could be further exploited. For example, the data could be exploited to cause economic issues to a target country (e.g., sudden restrictions on export of resources or medical supplies) or events intended to destabilize the country (e.g., sudden outbreak of diseases while a country's hospitals are nearing maximum capacity). The adversaries could cause further chaos if they successfully breach SCI systems and trigger anomalous conditions (e.g., opening/closing valves in critical infrastructure or overriding safety mechanisms) to cause emergency shutdowns, destruction of facilities or general chaos in transport systems.

Digital forensic investigators (DFI) and law enforcement agencies (LEA) have been crucial in investigating cyberattacks and cybercrime. However, DFI and LEA will likely require further support if they are called upon to investigate attacks on SCI as described earlier. This is because DFI and LEA personnel are typically not familiar of the unique characteristics 
of SCI systems as compared to conventional digital systems. Moreover, DFI and LEA are likely to have other ongoing digital investigations. Their ongoing case commitments may prevent them from being able to perform research on SCI systems to identify the required evidence. Meanwhile, SCI system owners may also be unable to provide the necessary evidence since these requirements were not specified in the first place when SCI was implemented. Caviglione et al. ~\cite{Caviglione_2017} stressed that digital forensic investigation on IoT devices would have to be performed on small-scale implementations (such as homes) to large-scale deployments such as those within a smart city. It was further stated that the technology used to build the infrastructure would be diverse and that their accessibility features must be discerned and appraised independently for forensic investigations~\cite{Caviglione_2017}. The predicament has not gone unnoticed - there have been calls for further digital forensic research on new digital artifacts and preventing misinterpretations of artifacts~\cite{HORSMAN2019100003}.

Digital forensics and cybersecurity in smart cities had been previously discussed, where the definition of smart cities was based on National Institute of Standards and Technology (NIST) and only focused on smart environments, living and mobility~\cite{BaigZubair2017}. While the threats, forensic data, and data sources highlighted did provide some guidance for DFI and LEA~\cite{BaigZubair2017}, the provided information cannot apply to all SCI internationally. For example, different countries have different technical requirements, implementation and data needs for their SCI. This could prove problematic for solutions that suggest a one-size-fits-all approach by specifically listing functional systems, such as the ones suggested by Baig et al.~\cite{BaigZubair2017}.

Due to the complexity of SCI, digital evidence for various components of SCI must be identified before actual cybercrime occurs. This is to reduce the stress faced by DFI and LEA as first responders to cybercrime. DFI and LEA could also better handle SCI cybercrime challenges if a standard definition of SCI is achieved, along with potential threats, offences and evidence sources pre-identified. Although this is a global challenge, adhering to global initiatives and standards allow flexibility of adoption by international DFI and LEA.


The contributions of our research are summarized as follows:
\begin{enumerate}
    \item We highlight current issues in SCI and define a standardized definition of SCI. 
    \item We develop and make publicly available our threat model template to governments, DFI and LEA to identify threats in SCI. 
    \item We map SCI threats to possible offences and corresponding SCI evidence sources and types. 
    \item We discuss future SCI digital forensics opportunities from a technical and legal perspective.
\end{enumerate}

For reproducibility and advancing the research in SCI digital forensics and threat modeling, our threat model is publicly available at: \url{https://github.com/poppopretn/SmartCityThreatModel}{.} 

The rest of this paper is organized as follows. In Section~\ref{Background}, we present the context of the paper and define SCI. In Section~\ref{SmartCityThreatModel}, we highlight the choice of our threat modeling methodology, showcase our threat model and present the threats identified in SCI. In Section~\ref{SmartCityCyberCrime}, we show the threats, offences, evidence sources and types within SCI that we derived using our threat model. In Section~\ref{DigitalForensics}, we discuss future technical and legal opportunities for SCI digital forensics. In Section~\ref{Limitations}, we explain the limitations of our research. In Section~\ref{RelatedWork}, we summarize current related work in SCI digital forensics. Finally, we conclude the paper in Section~\ref{Conclusion}. 
\section{Contextualizing Smart City Infrastructure} \label{Background}

Technological innovations and rapid deployment of Internet of Things (IoT) devices have transformed many cities in different geographical regions into smart cities~\cite{Al-Turjman2019}. Arguably, these cities may not have implemented sufficient infrastructure that can deliver futuristic societal outcomes such as accident-free environments or zero-waste scenarios. However, these current implementations have brought about positive changes such as moulding the design of future cities and achieving sustainable use of resources~\cite{Al-Turjman2019}. 

\subsection{Current Issues in Smart City Infrastructure} \label{SmartCityInfraIssues}

The implementation of SCI is an attractive option for 
governments looking to improve citizens' lives and has increased 
visibility to critical indicators such as resource utilization and 
public safety. Nonetheless, there are 
multiple challenges to implement such capabilities as outlined below:

\begin{enumerate}
\item \textbf{Definition Issues:} It is vital to set the right context and definition when SCI is discussed. From the academic perspective, a commonly agreed definition of a Smart City has yet to be agreed on. A brief literature review of papers regarding SCI was conducted and yielded at least three different definitions of a Smart City~\cite{Giffinger2007, ALTURJMAN2018327, Al-TurjmanFadi20205Ufm}.
	
SCI is not clearly defined from the industry perspective either. Various industry solutions such as Bosch~\cite{Bosch}, Cisco Kinetic for Cities~\cite{Cisco}, Microsoft CityNext~\cite{Microsoft} and Schneider Electric EcoStruxure~\cite{Schneider} have offered products touted to allow prospective customers to create smart cities. However, a review of their respective product briefs showed that these solutions appear not to be based on any commonly agreed upon definitions or standards. Many also fail to realize that such forms of definition are constrained by financial budgets and technological maturity of the location SCI are deployed in. This inevitably contributes to the problem of a lack of standard definition in SCI.
	
\item \textbf{Interoperability Issues:} This is an extension of the definition issue mentioned previously. There were at least 31 different vendors~\cite{ABIResearch} offering various platforms and technologies to build SCI. It is unlikely that any one vendor could meet all the design requirements (hardware and software) of a city/country. A more realistic outcome would be a myriad of vendors being chosen to implement technologies by their respective strengths. With such a gamut of sensors, protocols, and technologies, interoperability between vendors becomes an issue. Although entities such as FIWARE~\cite{FIWARE} and the TALQ Consortium~\cite{TALQ} offer Application Programming Interfaces (APIs) to allow interoperability of technologies with different vendors, adoption and implementation of such APIs remain unclear. 
	
\item \textbf{Cybercrime Issues:} Cyber attacks on smart cities could become the next issue governments face as such projects are implemented. From the legal perspective, respective laws possibly were not updated to include attacks on SCI. Meanwhile, from the law enforcement and incident response perspective, there could be a lack of experience, knowledge and training for professionals called upon to investigate such attacks. This issue is further exacerbated by the definition and interoperability issues. There is no standard definition of a smart city, and multiple technologies are being utilized in a smart city. Legal, remediation and law enforcement actions are hampered due to varying understanding of SCI and technology complexities, leading to a risk of misleading evidence being retrieved and presented to courts of law.
\end{enumerate}

A properly defined and widely accepted definition of SCI could address the issues highlighted above. For example, a properly defined SCI will facilitate and empower Digital Forensic Investigators (DFI) in peer review processes, especially at Levels 3 and 4 of the Peer Review Hierarchy as proposed by~\cite{HORSMAN2020301062}. It also facilitates interoperability between vendors and enhances the development of APIs. Finally, it enhances clarity in connections between evidence and criminal hypotheses, reducing the risks of misleading evidence being presented in courts~\cite{SMIT2018128}. 

\subsection{Defining Smart City Infrastructure} \label{SmartCityDefinition}

The primary issue originates from a lack of a standard definition in SCI as various vendors and entities are vying to be \textit{the} standard for smart cities. Geographical differences and individual governmental requirements have not helped foster a standard definition of a smart city. A way to transcend such challenges in defining a smart city, along with relevant data indicators is required to facilitate the resolution of issues raised in Section~\ref{SmartCityInfraIssues}. 

\begin{figure}[ht]
\includegraphics[width=0.9\columnwidth]{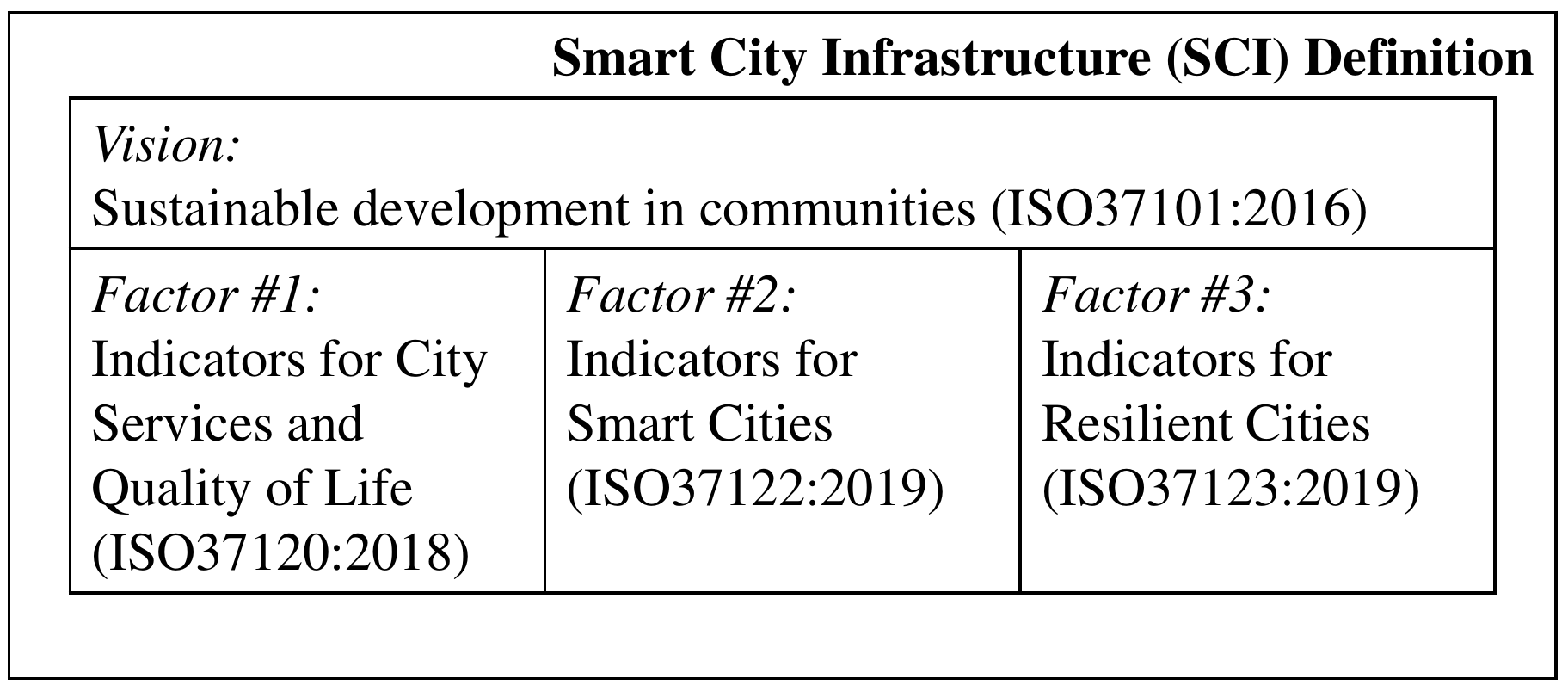}
\centering
\vspace{-5pt}\caption{Defining SCI}\vspace{-18pt}
\label{fig:2_background_SCI_Defn}
\end{figure}

As a body that strives to standardize methods to accomplish a final goal, the International Organization for Standardization (ISO) facilitates such an endeavour. After extensive research, we identified multiple ISO standards that provided a suitable framework for a standardized definition of SCI. With reference to~\autoref{fig:2_background_SCI_Defn}, the ISO standards that were selected are as follows:

\begin{enumerate}
	\item ISO37101:2016~\cite{ISO_37101_2016}
	
	\item ISO37120:2018~\cite{ISO_37120_2018}
	
	\item ISO37122:2019~\cite{ISO_37122_2019}
	
	\item ISO37123:2019~\cite{ISO_37123_2019}
\end{enumerate}

As observed from~\autoref{fig:2_background_SCI_Defn}, our proposed SCI definition (i.e. the body) is guided by four ISO standards. ISO37101:2016 serves as the guiding vision of a Smart City 
(i.e., a skeleton) via sustainable development. Concurrently, corresponding core and supporting data indicators from Factor \#1 (ISO37120:2018), Factor \#2 (ISO37122:2019) and Factor \#3 (ISO37123:2019) provide the required context of the SCI 
(i.e., the muscles/flesh).  Core data indicators are mandatory data indicators that must be captured if the ISO standards are used, whereas supporting data indicators are recommended to be captured (but not mandatory). We further explain our choice as follows:
\begin{enumerate}

\item \textbf{ISO37101:2016} - In contrast to a technical view and definition of SCI, the standard ISO37101:2016 adopts a technology-agnostic approach and uses sustainable development as a common denominator before any form of technical implementation is utilized. \autoref{table:1} lists the underlying fundamental purposes of sustainability in modern society along with sustainability issues raised in ISO37101:2016. Since the issues are systematically listed out before any form of smart city technology is implemented, ISO37101:2016 serves as a suitable component to drive the vision aspect of our proposed SCI definition (see~\autoref{fig:2_background_SCI_Defn}). 

\begin{table}[H]
\caption{Sustainability Purposes and Issues Highlighted in ISO37101:2016}
\begin{tabu} to 0.9 \columnwidth { | m{3.9cm} | m{4cm}| } 
 \hline
 \hfil\textbf{Purposes of Sustainability} & \hfil\textbf{Sustainability Issues} \\
  \hline
\vspace*{6pt}1. Attractiveness (e.g., Culture, identity) \newline 2. Preservation and improvement of environment (e.g., Protection of biological diversity and ecosystem) \newline 3. Resilience (e.g., Climate change adaptation, economic shock preparedness) \newline 4. Responsible resource use (e.g., Sustainable production, reusing and recycling of materials) \newline 5. Social cohesion (e.g., Diversity, sense of belonging, social mobility) \newline 6. Well-being (e.g., Happiness, healthy environment, quality of life) & 1. Governance, empowerment and engagement \newline 2. Education and capacity building \newline 3. Innovation, creativity and research \newline 4. Health and care in the community \newline 5. Culture and community identity \newline 6. Living together, interdependence and mutuality \newline 7. Economy and sustainable production and consumption \newline 8. Living and working environment \newline 9. Safety and security \newline 10. Community infrastructures \newline 11. Mobility \newline 12. Biodiversity and ecosystem services \\
\hline
\end{tabu}
\label{table:1}
\end{table}

\item \textbf{ISO37120:2018, ISO 37122:2019 and ISO37123:2019} - Building on the technology-agnostic approach of ISO37101:2016, core and supporting data indicators with respect to measuring city services and quality of life (ISO37120:2018), smart cities (ISO37122:2019) and resiliency (ISO37123:2019) are outlined. The data indicators used in the three ISO standards are listed in~\autoref{table:2}. While it is tempting to solely focus on data indicators listed in ISO37122:2019 instead of also using ISO37120:2018 and ISO37123:2019, the standard ISO37120:2018 has stated that ISO37122:2019 and ISO37123:2019 have to be used in conjunction with ISO37120:2018~\cite{ISO_37120_2018}. Further examination of ISO37122:2019 and ISO37123:2019 also yielded the same considerations that the three ISO standards must be used in tandem to provide a complete overview of a modern city~\cite{ISO_37122_2019, ISO_37123_2019}. Finally, another added advantage is that the data indicators in these three ISO standards can also be mapped back to the 17 United Nations Sustainable Development Goals (SDGs) 2015~\cite{ ISO_37120_2018, ISO_37122_2019, ISO_37123_2019, UNSDG}. This would allow governments to have better visibility in their progress towards the United Nations SDGs.
	
\begin{table}[H]
\caption{Data Indicators Listed in ISO37120:2018, ISO37122:2019 and ISO37123:2019}
\centering
\begin{tabu} to 0.9 \columnwidth { | M{2cm} | m{5cm}| } 
 \hline
 \textbf{Clause Number} & \hfil\textbf{Data Indicators} \\
 \hline
 5 & Economy \\
 \hline
 6  & Education  \\
 \hline
 7  & Energy  \\
 \hline
 8  & Environment and Climate Change\\
 \hline
 9  & Finance  \\
 \hline
 10  & Governance \\
 \hline
 11 & Health \\
 \hline
 12 & Housing  \\
 \hline
 13 & Population and Social Conditions  \\
 \hline
 14 & Recreation  \\
 \hline
 15 & Safety  \\
 \hline
 16 & Solid Waste  \\
 \hline
 17 & Sport and Culture  \\
 \hline
 18 & Telecommunication  \\
 \hline
 19 & Transportation  \\
 \hline
 20 & Urban/Local Agriculture and Food Security  \\
 \hline
 21 & Urban Planning  \\
 \hline
 22 & Wastewater  \\
 \hline
 23 & Water  \\
 \hline
\end{tabu}
\label{table:2}
\end{table}
	
\end{enumerate}	

Based on the information provided in the preceding paragraphs, we formally define SCI as follows:

\begin{definition} \label{SCI_Definition}
Smart City Infrastructure (SCI) are infrastructure designed to fulfil and address the six purposes of sustainability and twelve sustainability issues (based on ISO37101:2016 and outlined in~\autoref{table:1}), with technical implementations that provide visibility to data indicators set out in ISO37120:2018, ISO37122:2019 and ISO37123:2019 (outlined in~\autoref{table:2}). 
\end{definition}
\section{Threat Modeling Smart City Systems} \label{SmartCityThreatModel}

New systems likely have to be designed and built to capture the data indicators based on our proposed definition of SCI (\textit{i.e.,} Definition~\ref{SCI_Definition}). There could be multiple opportunities where cybersecurity threats creep into the designed systems despite endeavours to reduce such threats. Hence, it is imperative to ensure such systems are appropriately classified and devise a concrete means to identify potential threats. An appropriate Threat Modeling methodology should also be selected, with a means of clearly documenting identified threats to facilitate threat mitigation efforts. 

\subsection{Classification of Smart City Infrastructure Systems} \label{SmartCityInfraClassification}
\begin{figure}[ht]
\includegraphics[width=\columnwidth]{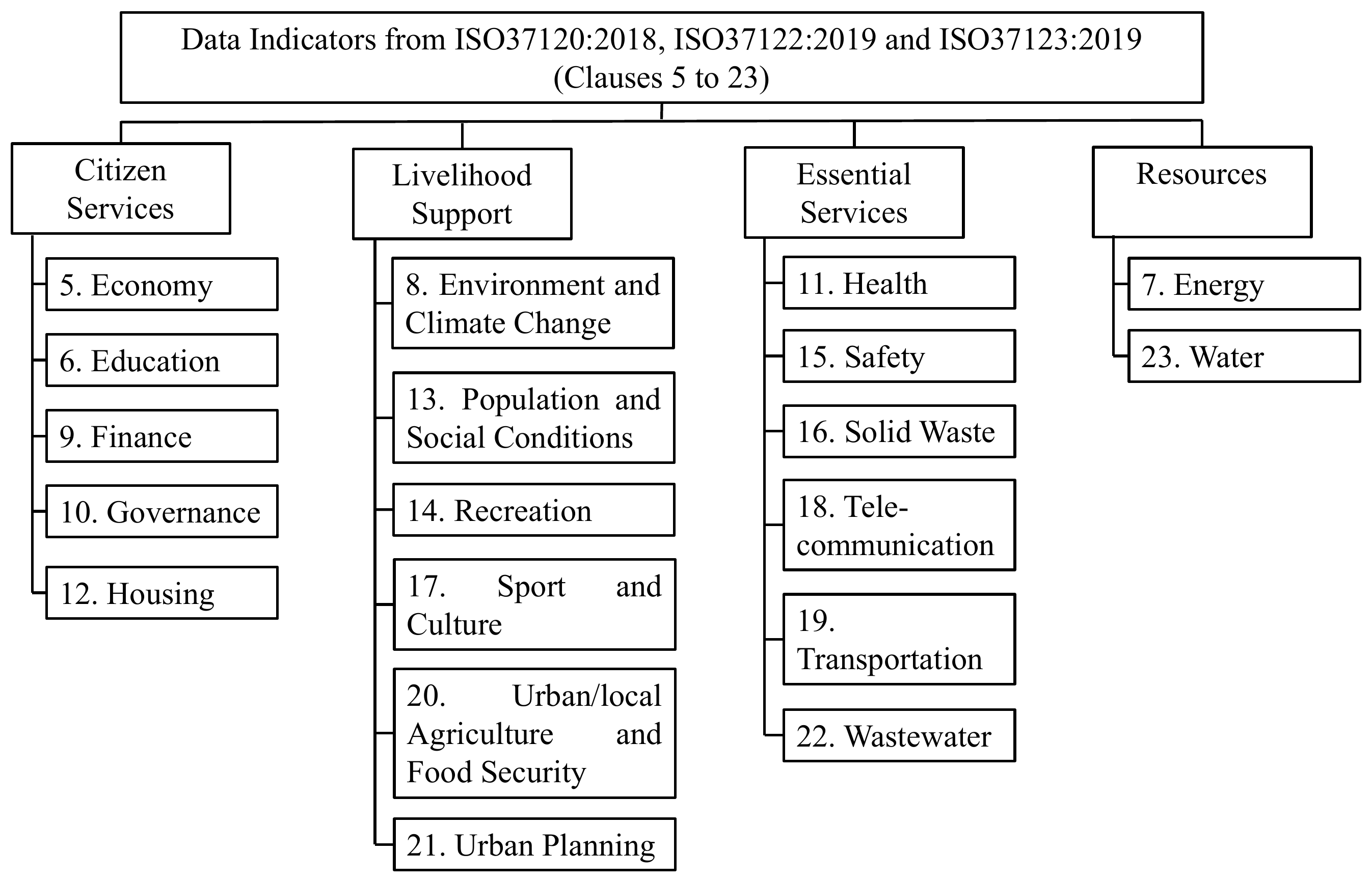}
\centering
\vspace{-15pt}\caption{Smart City Data Indicators Classification}\vspace{-5pt}
\label{fig:3_smartcitytm_system_classification}
\end{figure}

Multiple data indicators (Clauses 5 to 23) shown in~\autoref{table:2} appear to be daunting at first~\cite{ ISO_37120_2018, ISO_37122_2019, ISO_37123_2019}. However, after careful analysis, we were able to further classify the data indicators based on their respective outcomes into 4 distinct groups. With reference to~\autoref{fig:3_smartcitytm_system_classification}, the 4 groups are \textbf{\textit{Citizen Services}}, \textbf{\textit{Livelihood Support}}, \textbf{\textit{Essential Services}} and \textbf{\textit{Resources}}. These identified groups were used as a foundation for our proposed Smart City infrastructure system design and threat model.

\subsection{Selection of Threat Modeling Methodology} \label{ThreatModelingMethodSelection}

Threat modeling is defined as using abstractions to assist in the identification of risks~\cite{Shostack_TM}. It is best executed when systems are still in the design phase since identified risks can be addressed by modifying the design of the systems before they are formally implemented. Ideally, it should be executed after any proposed design change to ensure that potential threats are not introduced due to the change in the system. Having said that, threat modeling is still useful and important for systems that have deployed as well. However, it is likely that additional efforts and resources would have to be utilized to address the risks identified. This is because such systems would have been operational. For example, if the rectification of risks identified in an operational system requires it to be switched off, the system owners would have to divert operational requirements to another asset. This may, in turn, introduce complexities and potential risks. As such, it is highly encouraged that threat modeling should be done before any system is formally deployed.

Given a wide range of threats faced by various systems, multiple different threat modeling methodologies have been proposed throughout the years to identify threats in different scenarios. A collection of threat modeling methodologies have been identified and documented by Shevchenko et al.~\cite{CMU_SEI_2018} and presented as follows:

\begin{enumerate}
    \item \textbf{STRIDE: }Spoofing, Tampering, Repudiation, Information Disclosure, Denial of Service, Elevation of Privilege (STRIDE) was first formulated in 1999 by Loren Kohnfelder and Praerit Garg~\cite{MSFT_Adam_Shostack}. STRIDE has been successfully used in modeling threats in Cyber-Physical Systems~\cite{Shostack_TM}, and a software modeling tool is also available~\cite{MSFT_TMT}. 
    
    \item \textbf{PASTA: } Process for Attack Simulation and Threat Analysis (PASTA) was developed in 2012 and has seven stages of modeling activities. It requires both technical and non-technical stakeholders to participate in the process~\cite{PASTA_2015}.
    
    \item \textbf{LINDDUN: }Linkability, Identifiability, Non-Repudiation, Detectability, Disclosure of Information, Unawareness, Non-Compliance (LINDDUN) is a six-step process that is focused on modeling privacy-related threats. A noticeable drawback of this method lies in its complexity and labor-intensive process~\cite{CMU_SEI_2018}.
    
    \item \textbf{Attack Trees: } This process was originally developed by Bruce Schneider in 1999~\cite{AT_Schneier} and generally requires highly skilled individuals in cybersecurity to be involved in the process. It is a manual process without any guidelines for attack/risk identification, though some attempts to automate the process has been explored~\cite{Vigo_2014}.
    
    \item \textbf{OCTAVE: }Operationally Critical Threat, Asset, and Vulnerability Evaluation (OCTAVE) is a risk-based and strategic approach that consists of three phases~\cite{OCTAVE_2003}. It takes a different approach to threat modeling and identifies organizational risks instead of technical risks. 
    
    \item \textbf{VAST: }Visual, Agile, and Simple Threat (VAST) Modeling is a threat modeling strategy utilized by a commercial tool named ThreatModeler~\cite{VAST_2018}. VAST uses application and operational threat models - Process Flow Diagrams (PFDs) are created for application threat models, and operational threat models are built from analyzing the PFDs. Although VAST is useful, detailed specifications of VAST are not available since it is a commercial product.
    
    \item \textbf{CVSS: }Common Vulnerability Scoring System (CVSS) is a method that assigns a numerical score to a vulnerability based on its severity~\cite{CVSS_2019}. Widely used to determine the severity of vulnerabilities, it is not suitable to be used on its own in SCI as it would only display potential vulnerabilities in a numerical format. CVSS is often used in conjunction with other threat modeling methods~\cite{CMU_SEI_2018}.
    
    \item \textbf{Trike: }Introduced in 2006, Trike is a threat modeling method that could generate threats and attack trees automatically~\cite{Trike_2006}. However, the methodology and tools have not been updated for a long time and may not be suitable for modern SCI.
    
    \item \textbf{Persona non Grata: } Persona non Grata is a threat modeling method that focuses on the human aspects of threats~\cite{PnG_2014}. Although it has low false positives, it only identifies the human aspects of threats. 
    
    \item \textbf{Security Cards: } Developed in 2013, Security Cards is geared towards brainstorming about non standard situations and rarely used in the industry~\cite{CMU_SEI_2018}.
    
    \item \textbf{Quantitative Threat Modeling Method: }Developed in 2016 and synergizing STRIDE, Attack Trees and CVSS, the creators aimed to address issues arising from complex interdependence in components~\cite{QTMM_2016}. Unfortunately, while this method appears to be comprehensive and structured, the tool mentioned in the paper was not released publicly for usage.
    
    \item \textbf{hTMM: }Hybrid Threat Modeling Method (hTMM) combines threat modeling methodologies such as STRIDE, Persona non Grata and Security Cards for a focused threat modeling activity~\cite{hTMM_2018}. However, the process is still largely manual and lacks a dedicated tool that can execute the entire process of hTMM.
\end{enumerate}

\begin{table}[H]
\begin{center}
\centering
\caption{Comparison of Various Threat Modeling Methodologies}
\resizebox{\columnwidth}{!}
{\begin{tabu}to 0.9 \columnwidth{| M{2cm} | m{1.6cm} | m{1.6cm} | m{1.6cm}|} 
    \cline{1-4}
    Threat Modeling Methodology  & Software Tool Support & Ongoing Updates & \vspace*{3pt}Suitability for Smart City TM\vspace*{3pt}  \\
    \hline
    \vspace*{3pt}STRIDE\vspace*{3pt} & \vfil\hfil\cmark & \vfil\hfil\cmark & \vfil\hfil\cmark \\
    \hline
    \vspace*{3pt}PASTA\vspace*{3pt} & \vfil\hfil\xmark & \vfil\hfil\xmark & \vfil\hfil\cmark \\
    \hline
    \vspace*{3pt}LINDDUN\vspace*{3pt} & \vfil\hfil\xmark & \vfil\hfil\xmark & \vfil\hfil\cmark \\
    \hline
    \vspace*{3pt}Attack Trees\vspace*{3pt} & \vfil\hfil\cmark & \vfil\hfil\xmark & \vfil\hfil\xmark \\
    \hline    
    \vspace*{3pt}OCTAVE\vspace*{3pt} & \vfil\hfil\xmark & \vfil\hfil\xmark & \vfil\hfil\xmark \\
    \hline    
    \vspace*{3pt}VAST Modeling\vspace*{3pt} & \vfil\hfil\cmark & \vfil\hfil\cmark & \vfil\hfil\cmark \\
    \hline    
    \vspace*{3pt}CVSS\vspace*{3pt} & \vfil\hfil\xmark & \vfil\hfil\xmark & \vfil\hfil\cmark \\
    \hline    
    \vspace*{3pt}Trike\vspace*{3pt} & \vfil\hfil\xmark & \vfil\hfil\xmark & \vfil\hfil\xmark \\
    \hline    
    \vspace*{3pt}Persona non Grata\vspace*{3pt} & \vfil\hfil\xmark & \vfil\hfil\xmark & \vfil\hfil\xmark \\
    \hline    
    \vspace*{3pt}Security Cards\vspace*{3pt} & \vfil\hfil\xmark & \vfil\hfil\xmark & \vfil\hfil\xmark \\
    \hline
    \vspace*{3pt}Quantitative Threat Modeling Method\vspace*{3pt} & \vfil\hfil\xmark & \vfil\hfil\xmark & \vfil\hfil\xmark \\
    \hline    
    \vspace*{3pt}hTMM\vspace*{3pt} & \vfil\hfil\xmark & \vfil\hfil\xmark & \vfil\hfil\xmark \\
    \hline
\end{tabu}
}
\label{table:3}
\end{center}
\end{table}

\subsubsection{Assessment Criteria for Threat Modeling Methodologies} \label{ThreatModelAssessmentCriteria}

We examined all twelve threat modeling methodologies to assess if there were any methodologies that would be a good fit for our research objectives. We believe that all the twelve threat modeling methodologies were worth examining as discussed by Shevchenko et al.~\cite{CMU_SEI_2018} accounts for the most popular and state-of-the-art threat modeling techniques. There were three key requirements - support for software tools based on the threat modeling methodology, ongoing updates and suitability for Smart City threat modeling.

These three essential requirements were identified as significant and relevant when determining the suitability of the chosen threat modeling methodologies. Firstly, a threat modeling methodology that is supported by a software tool allows collaborative work, enhances design efficiency and reduces potential errors when designing a threat model (e.g. in-built notations and feedback notifications). Secondly, a constantly updated threat model methodology ensures that it remains relevant to the ever-changing cybersecurity landscape.  Finally, not all threat modeling methodologies may apply to the SCI context. As such, it is vital to identify which methodology is appropriate for SCI.

The details of our assessment are summarized in~\autoref{table:3}, and we selected STRIDE as our main Threat Modeling methodology for SCI systems designed based on Definition~\ref{SCI_Definition}. There were a multitude of reasons why STRIDE was selected. Firstly, a freely available and consistently updated threat modeling tool was available. As research into threats on SCI should be an ongoing process, the tool used for threat modeling should also be actively maintained. Secondly, the STRIDE threat modeling methodology pays close attention to threats in the systems while being less complex than other methodologies such as LINDDUN, PASTA and VAST. Although it was also argued that STRIDE could grow in complexity as the systems grow larger~\cite{CMU_SEI_2018}, the availability and usage of the Microsoft Threat Modeling Tool (TMT)~\cite{MSFT_TMT} could reduce the potential complexity. Finally, the Microsoft TMT is freely available for all, while methodologies such as VAST and tools such as IriusRisk~\cite{IriusRisk_2021} remain as commercial products. As costs associated with security can sometimes be an important factor, especially in novel systems, we aim to remove that barrier and foster the creation of secure systems during the design phase.

\subsection{Threat Modeling Smart City Infrastructure} \label{ThreatModelingSmartCity}

As explained in Section~\ref{ThreatModelingMethodSelection}, the STRIDE threat modeling methodology and the Microsoft TMT are used to create the corresponding threat model of the proposed Smart City System based on Definition~\ref{SCI_Definition}. A threat model designed with the Microsoft TMT can contain up to four layers, and are as follows~\cite{Microsoft_TM_DFD_Layer_2022}:

\begin{enumerate}
    \item \textbf{Layer 0: }This is known as the system layer (it can also be called the context layer), and is a compulsory layer in a threat model designed in Microsoft TMT as it is deemed as a starting point for any system~\cite{Microsoft_TM_DFD_Layer_0_2022}. Layer 0 contains core parts and processes of the system, and the corresponding data-flow diagrams should fit into a single page~\cite{Microsoft_TM_DFD_Layer_0_2022}.
    
    \item \textbf{Layer 1: }This is known as the process layer, and contains the secondary system parts that originate from Layer 0. Like the system layer, the diagram should also be contained within a single page. 
    
    \item \textbf{Layer 2: }This is known as the sub-process level, and contains details of the various components that the secondary systems are composed of. This layer is utilized when the corresponding systems handle sensitive data or are deemed as high-risk.
    
    \item \textbf{Layer 3: }This is known as the lower-level layer, and used when a kernel-level system is constructed. Every process and sub-process is described in detail via data-flow diagrams.
\end{enumerate}

\begin{figure}[htbp]
\includegraphics[width=\columnwidth]{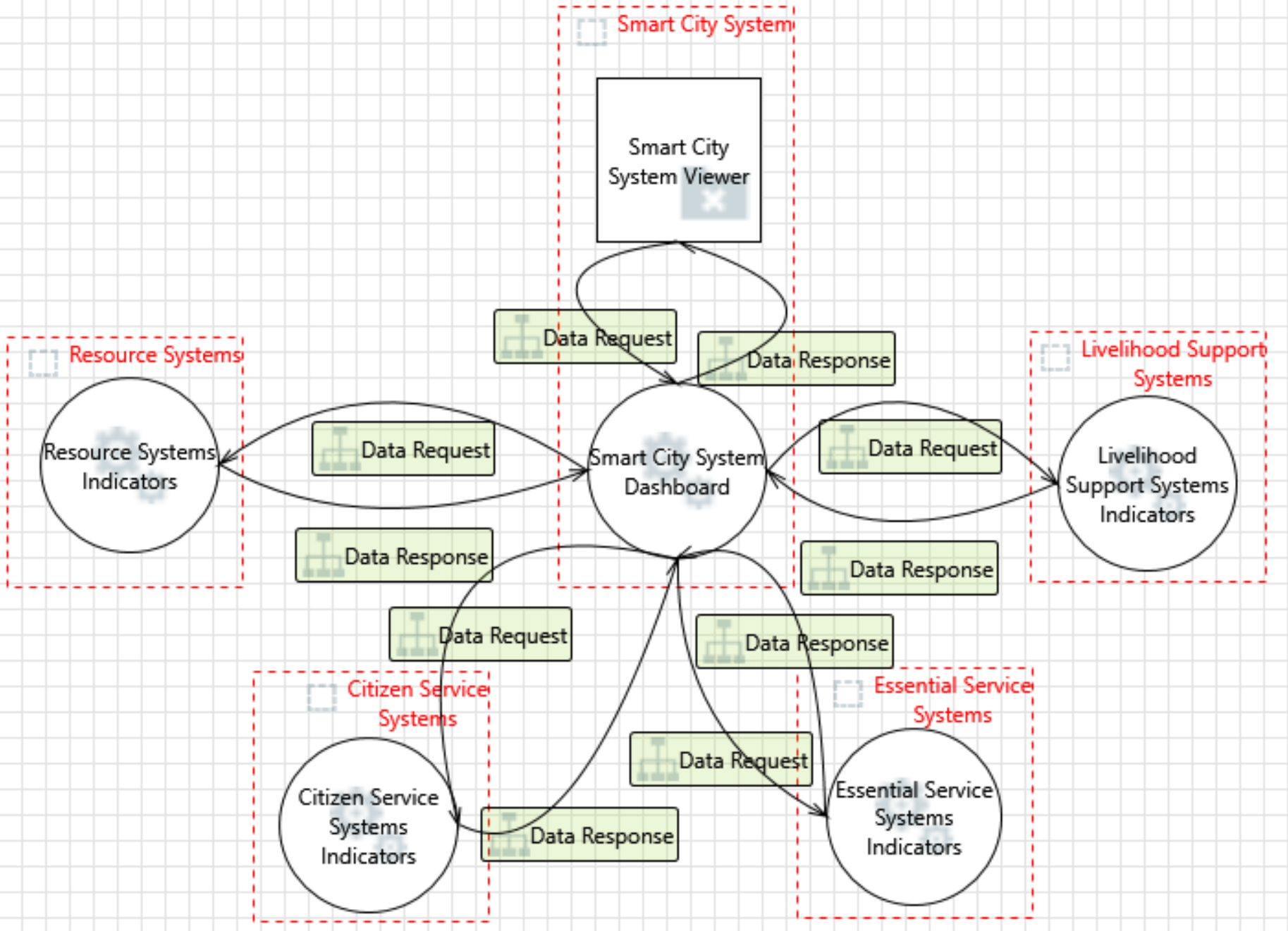}
\centering
\vspace{-15pt}\caption{Smart City System Threat Model - Layer 0}
\label{fig:3_smartcitytm_L0}
\end{figure}

\begin{figure}[htbp]
\includegraphics[width=\columnwidth]{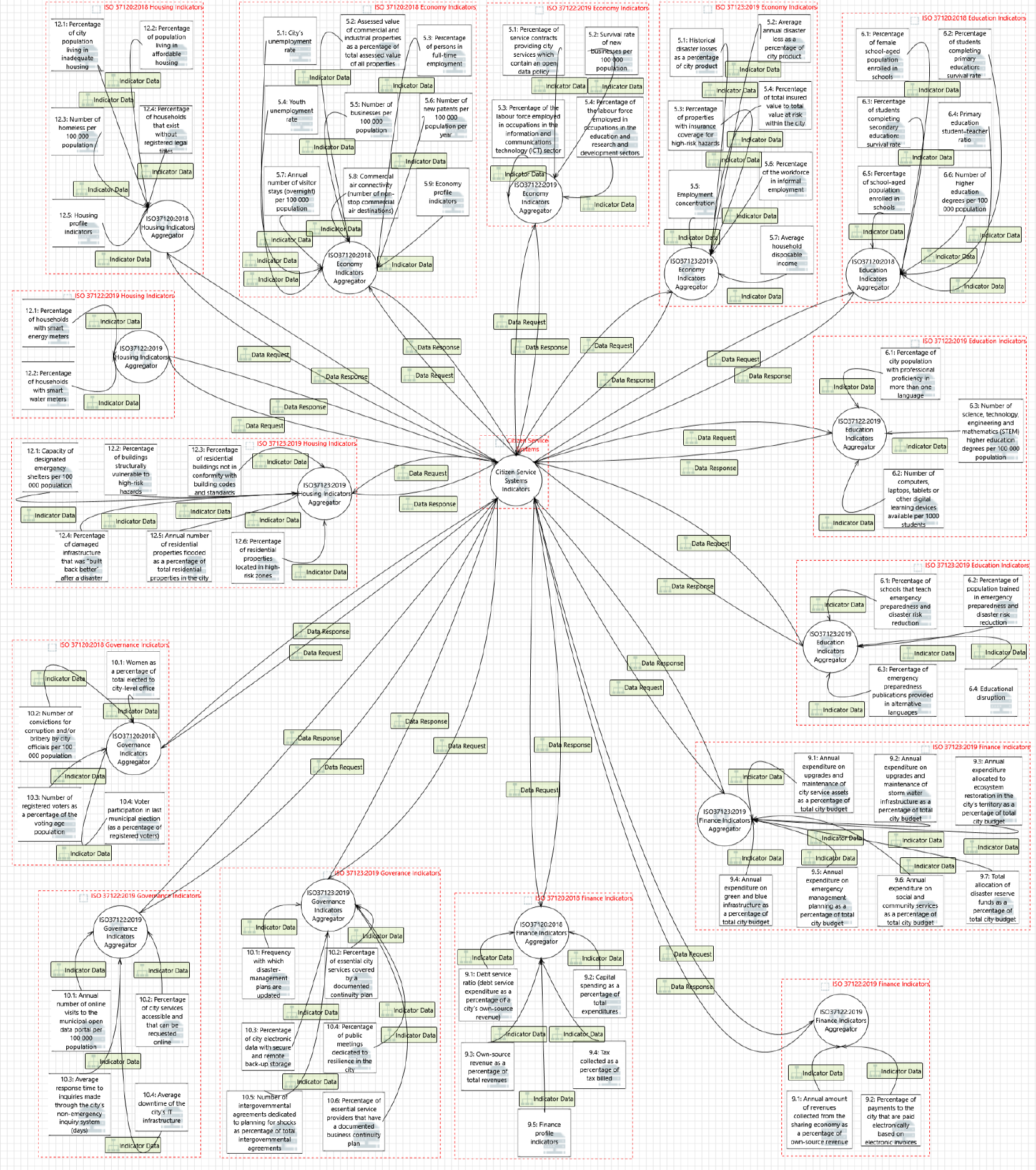}
\centering
\vspace{-15pt}\caption{Smart City Citizen Service System Threat Model - Layer 1}\vspace{-15pt}
\label{fig:3_smartcitytm_citizen_services_TM_L1}
\end{figure}

Based on our defined Smart City Infrastructure, we have deliberately designed the threat model to be up to Layer~1. Firstly, we envision our threat model to be used and applicable to the international audience and community. Different governments and countries may have additional requirements and technological preferences for their Smart City Infrastructure. If we had designed our threat model down to Layer 3, that would have been very specific. Being overly detailed and specific could result in our threat model being overlooked instead. This is because it would require further efforts to review and amend the threat model. We have assessed that designing our threat model up to Layer 1 offered the best baseline - It was not overly simplistic and contained the necessary core and secondary systems required to capture the data defined by Definition~\ref{SCI_Definition}. It also sets the stage for potential users to customize their desired Layers 2 and 3 threat models (if applicable). With reference to~\autoref{fig:3_smartcitytm_L0} and~\autoref{fig:3_smartcitytm_citizen_services_TM_L1}, a screenshot of Layer 0 and Layer 1 of the Citizen Service System (one of the secondary systems) is shown. 

In the next section, we discuss the threats analyzed from 
our SCI threat models (\autoref{fig:3_smartcitytm_L0} and 
\autoref{fig:3_smartcitytm_citizen_services_TM_L1}).

\subsection{Threats in Smart City Infrastructure} \label{SmartCityInfraThreats}

\begin{figure}[htbp]
\includegraphics[width=0.8\columnwidth]{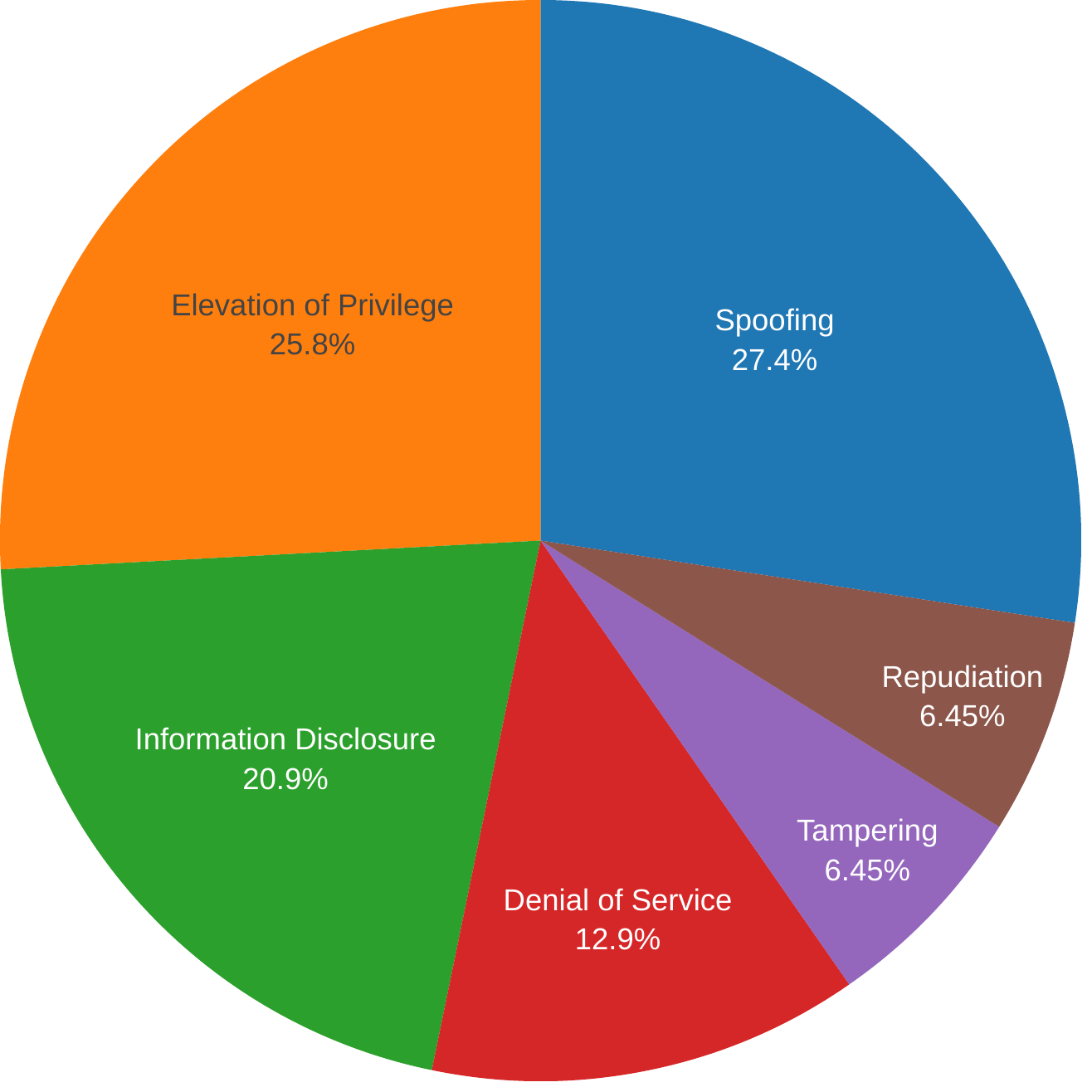}
\centering
\caption{Overview of Threats Derived from Threat Model Grouped by Category}
\label{fig:3_Threat_Overview_Pie_Chart_STRIDE}
\end{figure}

The threat modeling activity in Section~\ref{ThreatModelingSmartCity} yielded interesting results. There were a total of 1768 potential threats across the six threat categories of STRIDE. The total number and types of each threat category are summarized in~\autoref{table:4} and~\autoref{fig:3_Threat_Overview_Pie_Chart_STRIDE}, with threats from the Elevation of Privilege category having the greatest variety of threats.

\begin{table}[htbp]
\caption{Types of Threats Identified during Threat Modeling}
\begin{tabu} to 0.9\columnwidth  { | M{3cm} | m{5cm}| } 
 \hline
 \vspace*{6pt}\textbf{Category (Total Number of Threats)}\vspace*{6pt} & \hfil\textbf{Types of Threats}\\
 \hline
 Spoofing (485) & \vspace*{6pt}1. Spoofing in various processes \newline 2. Spoofing of various source data stores \vspace*{6pt}\\
 \hline
 Tampering (114)  & \vspace*{6pt}1. Potential lack of input validation \vspace*{6pt}\\
 \hline
 Repudiation (114)  & \vspace*{6pt}1. Potential data repudiation \vspace*{6pt}  \\
 \hline
 Information Disclosure (370)  & \vspace*{6pt}1. Data flow sniffing \newline 2. Potential weak access control for a resource \vspace*{6pt}\\
 \hline
 Denial of Service (228)  & \vspace*{6pt}1. Potential process crash or stop \newline 2. Data flow potentially interrupted \vspace*{6pt}\\
 \hline
 Elevation of Privilege (457)  & \vspace*{6pt}1. Systems may be subject to elevation of privilege using remote code execution \newline 2. Elevation using impersonation \newline 3. Cross site request forgery \newline 4. Elevation by changing execution flow in various systems \vspace*{6pt}\\
\hline
\end{tabu}
\label{table:4}
\end{table}

\begin{figure}[htbp]
\includegraphics[width=0.8\columnwidth]{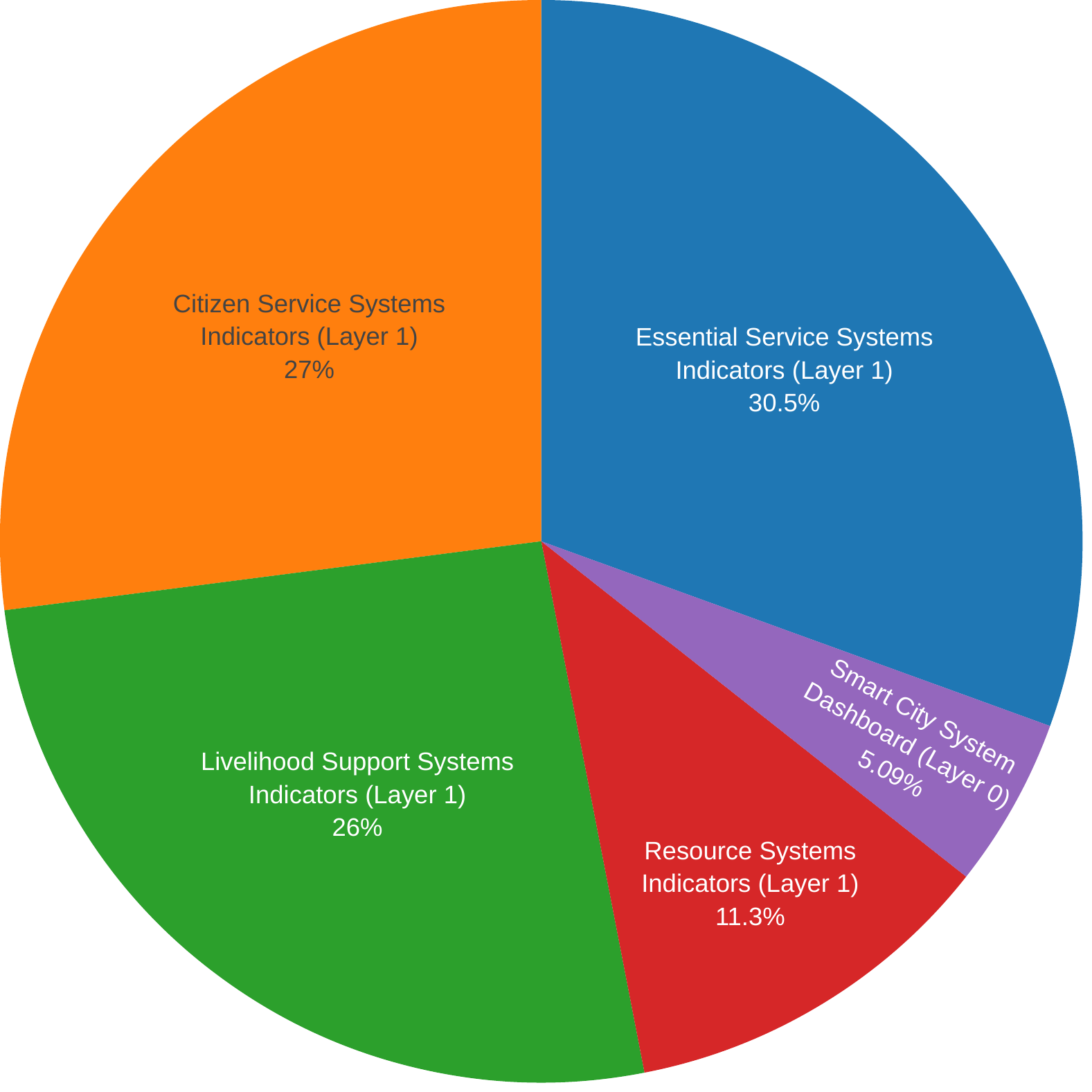}
\centering
\caption{Overview of Threats Derived from Threat Model}
\label{fig:3_Threat_Overview_Pie_Chart}
\end{figure}

~\autoref{fig:3_Threat_Overview_Pie_Chart} shows the breakdown of potential threats in our proposed SCI systems, with the Smart City System Dashboard (Level 0) and Resource Systems Indicators (Level 1) having a lower percentage of potential threats (5.09\% and 11.3\%, respectively). The smaller number of threats was expected as the Smart City System Dashboard only contained major system processes and was less complex since it was at Layer 0. Meanwhile, the Resource Systems Indicators had the smallest number of data indicators (as shown in ~\autoref{fig:3_smartcitytm_system_classification}). Thus, they had a smaller number of potential threats than the other Level 1 systems.

From the threat mitigation perspective, stakeholders can consider a few possible approaches based on the information presented in the preceding (i.e., \autoref{table:4}, ~\autoref{fig:3_Threat_Overview_Pie_Chart_STRIDE} and~\autoref{fig:3_Threat_Overview_Pie_Chart}). For a quick-win approach, stakeholders could consider mitigating threats related to the Tampering and Repudiation category and target the Smart City System Dashboard and Resource Systems Indicators first, as they are the lowest in number. If the stakeholders are interested in mitigating a system with the highest number of potential threats and its corresponding category,  the Essential Service Systems Indicators and Spoofing category threats could be reviewed. Ultimately, the threat model allows stakeholders to start studying or mitigating any threat they wish. Armed with the knowledge of potential threats that the SCI may face, defenders could also use these threats to anticipate potential cybercrime occurring on these systems.

In the next section, we leverage our SCI threat models 
to identify the cybercrimes on SCI systems.
\section{Cybercrime on Smart City Systems} \label{SmartCityCyberCrime}

Just like SCI, the term "cybercrime" could have differing interpretations. It is vital to iron out a few key terms that may cause confusion and set the scope of discussion. After defining cybercrime for SCI, we map threats to offences that could occur in SCI and, finally, the associated evidence classification and requirements.

\subsection{Defining Cybercrime in Smart City Systems} \label{SmartCityCyberCrimeDefinition}

Based on our literature review, we observed a few key terms that appeared to be used interchangeably, which could cause confusion and disagreement in our paper if not defined appropriately. The terms in question are "cybercrime", "computers", and "computer crime". The exact definition of cybercrime has been debated at length, especially with the evolution of the threat landscape over the years. Gillespie~\cite{GillespieAlisdair2019C:KI} asserts that it is a crime committed or facilitated by the Internet, with no distinguishing factor as to how access to the Internet was obtained. Meanwhile, with the advancement of technology, computers now come in various form factors and peripherals. Given the broad spectrum of hardware configuration, formally defining what a "computer" entails is difficult. The Budapest Convention on Cybercrime gave a broad definition and considered devices capable of transmitting data (whether in silo or networked) as a computer~\cite{Budapest_Convention_2001}. Finally, "cybercrime" and "computer crime" have often been used interchangeably to describe illegal activities on systems. Going by how the Budapest Convention on Cybercrime defines computers, "computer crime" should be defined as crimes carried out using computers. Since this paper uses the term "cybercrime", our scope would be limited to the crimes committed or facilitated by the Internet as described earlier. 

\subsection{Mapping Threats to Cybercrime and Offences}\label{MapThreatsToCybercrime}

Given the borderless nature of the Internet, cybercrime is inevitably an international issue that is further complicated by geopolitical boundaries. In an attempt to achieve some form of consensus between nations and various stakeholders, the Budapest Convention on Cybercrime (also known as the Budapest Convention) was formed on 23rd November 2001 by the Council of Europe~\cite{Budapest_Convention_2001}. As of April 2022, a total of 81 States are now either Parties (66), or have signed it or been invited to accede (15)~\cite{Budapest_Convention_2022}. This accounts for only about 42\% of the represented members in the United Nations (there are currently 193 members in the United Nations~\cite{United_Nations_2022}). Notable absentees from the signatories of the Budapest Convention include the Russian Federation and China (both permanent members of the United Nations Security Council). The absence of their support for the Budapest Convention could be due to geopolitical tensions and that they were not consulted when the Budapest Convention was first formulated, an observation that was also noted by Gillespie~\cite{GillespieAlisdair2019C:KI}. 

Unfortunately, there is yet a global consensus on cybercrime. Geopolitical issues, potential lack of awareness and urgency by international governments inadvertently facilitate cybercrime. However, there is a strong need for a collective agreement for cybercrime to be addressed primarily due to its potential global reach. This is especially so if various governments are working towards the implementation of SCI. Academic research, coupled with suitable theoretical models and adaptation of a sensible definition of cybercrime should help governments be more cognizant of the crimes that could occur.

\begin{table}[H]
\caption{Mapping Threats on SCI to Crimes and Offences}
{\begin{tabu} {|m{1cm}|m{2cm}|m{4.5cm}|}  
    \cline{1-3}
    \vspace*{6pt}\textbf{Type of Crime (based on~\cite{GillespieAlisdair2019C:KI})} \vspace*{6pt} & \textbf{Threats on SCI (based on~\autoref{table:4})} & \textbf{Possible Offences Committed (based on~\cite{Budapest_Convention_2001})}   \\
    \hline
    \multirow{12}{4em}{Crimes against computers} & \vspace*{6pt}Spoofing in various processes\vspace*{6pt} & Illegal Access (Article 2), Illegal Interception (Article 3), System Interference (Article 5), Misuse of Devices (Article 6) \\ 
    \cline{2-3}
    & Spoofing of various source data stores & \vspace*{6pt}Illegal Access (Article 2), Illegal Interception (Article 3), Data Interference (Article 4), System Interference (Article 5), Misuse of Devices (Article 6), Computer-related Forgery (Article 7) \vspace*{6pt}\\ 
    \cline{2-3}
    & \vspace*{6pt}Potential lack of input validation\vspace*{6pt} & Data Interference (Article 4), Misuse of Devices (Article 6), Computer-related Forgery (Article 7)\\ 
    \cline{2-3}
    & Potential data repudiation & \vspace*{6pt}Data Interference (Article 4), System Interference (Article 5), Misuse of Devices (Article 6), Computer-related Forgery (Article 7)\vspace*{6pt} \\ 
    \cline{2-3}
    & Data flow sniffing & \vspace*{6pt}Illegal Access (Article 2), Illegal Interception (Article 3), Misuse of Devices (Article 6) \vspace*{6pt}\\ 
    \cline{2-3}
    & \vspace*{6pt}Potential weak access control for a resource \vspace*{6pt} & Illegal Access (Article 2), Data Interference (Article 4), System Interference (Article 5), Misuse of Devices (Article 6) \\ 
    \cline{2-3}
    & Potential process crash or stop & \vspace*{6pt} Illegal Access (Article 2), Illegal Interception (Article 3), Data Interference (Article 4), System Interference (Article 5), Misuse of Devices (Article 6) \vspace*{6pt} \\ 
    \cline{2-3}
    & Data flow potentially interrupted & \vspace*{6pt}Illegal Access (Article 2), Illegal Interception (Article 3), Data Interference (Article 4), System Interference (Article 5), Misuse of Devices (Article 6) \vspace*{6pt}\\ 
    \cline{2-3}
    & \vspace*{6pt}Systems may be subject to elevation of privilege using remote code execution \vspace*{6pt}& Illegal Access (Article 2), Illegal Interception (Article 3), Misuse of Devices (Article 6)  \\ 
    \cline{2-3}
    & \vspace*{6pt}Elevation using impersonation \vspace*{6pt}& Illegal Access (Article 2), Illegal Interception (Article 3), Misuse of Devices (Article 6) \\ 
    \cline{2-3}
    & Cross site request forgery & \vspace*{6pt}Illegal Access (Article 2), Illegal Interception (Article 3), Misuse of Devices (Article 6) \vspace*{6pt} \\ 
    \cline{2-3}
    & \vspace*{6pt}Elevation by changing execution flow in various systems\vspace*{6pt} & Illegal Access (Article 2), Illegal Interception (Article 3), System Interference (Article 5), Misuse of Devices (Article 6), Computer-related Forgery (Article 7)  \\ 
    \hline
    \end{tabu}}
    \label{table:5}
\end{table}

After evaluating state-of-the-art works on cybercrime, we assessed that the adaptation of a subset of the taxonomy of cybercrime categories proposed by Gillespie~\cite{GillespieAlisdair2019C:KI} and a subset of the offences outlined by the Budapest Convention would be suitable to represent cybercrime in SCI. We mapped the types of threats outlined in~\autoref{table:4} to the corresponding cybercrime categories and offences, as shown in~\autoref{table:5} below.

There were other categories that Gillespie~\cite{GillespieAlisdair2019C:KI} proposed (e.g., crimes against property, illicit content and individuals). Similarly, the Budapest Convention had also included other types of offences e.g., child pornography, infringement of copyrights and abetting crimes. These are excluded from~\autoref{table:5} for being out of scope as we only focus on crimes on SCI.

\subsection{Evidence Requirements for SCI Threats and Offences} \label{EvidenceRequirements}

\begin{table*}[t]
\caption{Mapping Threats on SCI to Possible Evidence Types and Sources} \label{table:6}
\begin{adjustbox}{width=\textwidth,totalheight=\textheight,keepaspectratio}
{\begin{tabu}{|m{4.5cm}|m{14.7cm}|m{5.3cm}|}  
    \cline{1-3}
    \vspace*{6pt}\centering\textbf{Threats on SCI (based on~\autoref{table:4})} \vspace*{6pt} & \vspace*{6pt} \centering\textbf{Types of Possible Evidence \vspace*{6pt}} & \vspace*{6pt} \centering \textbf{Possible Evidence Sources}  \vspace*{6pt}  \\
    \hline
    \vspace*{6pt}Spoofing in various processes\vspace*{6pt} & \multirow{4}{*}{\parbox{14cm}{\vspace*{25pt} 1. Network traffic (captured from SCI systems), device logs, server logs, system logs, malicious binaries  \newline 2. Network traffic (captured from SCI network infrastructure) \newline 3. Memory images, hard disk images, devices used to commit crimes \vspace*{6pt}}} &  \multirow{9}{*}{\parbox{14cm}{\vspace*{80pt} 1. SCI systems \newline 2. SCI network infrastructure  \newline 3. Adversarial systems \vspace*{6pt}}}    \\ \cline{1-1}
    \vspace*{6pt} Potential process crash or stop \vspace*{6pt} &  &  \\ 
    \cline{1-1}
    \vspace*{6pt}Data flow potentially interrupted\vspace*{6pt} &   &  \\ 
    \cline{1-1}
    \vspace*{6pt}Elevation by changing execution flow in various systems\vspace*{6pt} & \ &  \\ 
    \cline{1-2}
    \vspace*{6pt}Spoofing of various source data stores\vspace*{6pt} & \vspace*{6pt} 1. Network traffic (captured from SCI systems), device logs, server logs, system logs, SCI data stores  \newline 2. Network traffic (captured from SCI network infrastructure) \newline 3. Memory images, hard disk images, devices used to commit crimes \vspace*{6pt} & \\ 
    \cline{1-2}
    \vspace*{6pt}Potential lack of input validation\vspace*{6pt} & \multirow{2}{*}{\parbox{14cm}{\vspace*{8pt} 1. Network traffic (captured from SCI systems), device logs, server logs, system logs  \newline 2. Network traffic (captured from SCI network infrastructure) \newline 3. Memory images, hard disk images, devices used to commit crimes \vspace*{8pt}}}  &  \\ 
    \cline{1-1}
    \vspace*{6pt}Potential weak access control for a resource\vspace*{6pt} &  &  \\ 
    \cline{1-2}
    \vspace*{6pt}Elevation using impersonation \vspace*{6pt} & \multirow{2}{*}{\parbox{14cm}{\vspace*{10pt} 1. Network traffic (captured from SCI systems), device logs, server logs, system logs, malicious binaries, SCI Executable programs  \newline 2. Network traffic (captured from SCI network infrastructure) \newline 3. Memory images, hard disk images, devices used to commit crimes \vspace*{6pt}\vspace*{6pt}}} &  \\ 
    \cline{1-1}
    \vspace*{6pt}\vspace*{6pt}Systems may be subject to elevation of privilege using remote code execution \vspace*{6pt} &   &  \\ 
    \cline{1-3}
    \vspace*{6pt}Potential data repudiation \vspace*{6pt} & \vspace*{6pt} 1. Network traffic (captured from SCI systems), device logs, server logs, system logs \newline 2. Network traffic (captured from SCI network infrastructure) \vspace*{6pt} & \vspace*{10pt} 1. SCI systems \newline 2. SCI network infrastructure \vspace*{6pt}\\ 
    \cline{1-3}
    \vspace*{6pt}Data flow sniffing\vspace*{6pt} & \vspace*{6pt} 1. Device logs, server logs, system logs, malicious binaries \newline 2. Devices used to commit crimes, hard disk images, memory images \vspace*{6pt} & \vspace*{6pt} 1. SCI systems \newline 2. Adversarial systems \vspace*{6pt} \\ 
    \cline{1-3}
    \vspace*{6pt}Cross site request forgery\vspace*{6pt} & \vspace*{6pt} 1. Network traffic (captured from SCI systems), web server logs  \newline 2. Network traffic (captured from SCI network infrastructure) \newline 3. Browser history, device artifacts (e.g. registry keys for Windows systems) \newline 4. Memory images, hard disk images, devices used to commit crimes \vspace*{6pt}  & \vspace*{6pt} 1. SCI web servers \newline 2. SCI network infrastructure \newline 3. User devices \newline 4. Adversarial systems \vspace*{6pt}  \\ 
    \hline
    \end{tabu}}
    \end{adjustbox}
\end{table*}

Although the issue of pinpointing offences to threats on SCI has been addressed, it is necessary to identify the evidence required to prove such offences have occurred. Based on our earlier efforts in identifying types of threats during threat modeling SCI and the associated offences in~\autoref{table:5}, we further identified the possible evidence types that investigators could use. The evidence types were identified by referring to the threats on SCI and examining the corresponding possible evidence sources. After the evidence sources were determined, types of possible evidence (sorted in order of volatility~\cite{Casey2011}) linked to evidence sources were established. This is presented in~\autoref{table:6} below.

It can be observed that evidence sources and the various types of possible evidence are repeated in some threats on SCI. Although the types of possible evidence may be similar for the various threats on SCI, the context and content within the evidence would be the critical difference for investigators. For example, the threat of spoofing in various processes and potential process crash or stop are different in nature, but the types of possible evidence to distinguish the threat are the same (e.g., network traffic and memory images). 
\section{Digital Forensics Implications and Opportunities} \label{DigitalForensics}

The earlier sections of the paper defined SCI, identified SCI threats, offences committed and the corresponding evidence that could prove such offences were committed. However, it can be argued that SCI is still a work in progress as an international implementation of SCI has yet to be completed. Although SCI has yet to be implemented globally, it presents a chance for law enforcement agencies and investigators to quickly seize the opportunity to develop their digital forensics capabilities. We first discuss how DFI and LEA could benefit from our research and then discuss the possible future opportunities in digital forensics. These are divided into 2 major categories - research and legal opportunities.

\subsection{Implications for Smart City Digital Forensics} \label{Implications_SC_Digital_Forensics}

We previously identified three issues that would hamper DFI and LEA in their investigation work on SCI attacks (namely definition, interoperability and cybercrime issues). Let us assume a theoretical scenario where a SCI has been fully implemented and consists of various technological implementations by multiple vendors. The successful implementation of the SCI draws the attention of cybercriminals and sophisticated state-sponsored adversaries, who commence their cyber offensive operations on SCI (e.g. energy monitoring systems and the associated sensors of the nation's households). DFI and LEA are called upon to investigate the multitude of cyber-attacks. Still, they have to contend with current investigative workloads, understand the nuances of SCI, and various underlying technologies and finally obtain the proper evidence to determine the offences that had taken place. In this instance, DFI and LEA could expect to take a long time to investigate the attacks while also trying to resolve their current workload.

However, with our contributions, DFI and LEA can mitigate many challenges. Assuming that the SCI was designed using our threat model, the corresponding threats, cybercrime, and evidence required to identify the offences would have been documented. For example, based on~\autoref{table:5}, DFI and LEA would know that there are twelve types of threats within SCI. The adversaries caused a process crash or stop and might have exploited vulnerabilities to change execution flows in the energy monitoring systems and gained elevated privileges. The offences committed would be illegal access, illegal interception, data interference, system interference and misuse of devices for process crash or stop, while the offences committed for gaining elevated privileges via changing execution flow would be illegal access, illegal interception, system interference, misuse of devices and computer-related forgery. After knowing the offences the adversaries committed, DFI and LEA can retrieve the possible evidence sources based on the information presented in~\autoref{table:6}. After consulting~\autoref{table:6}, the investigators know that the evidence sources are the SCI systems, SCI network infrastructure and the systems used by adversaries to commit the crimes. DFI and LEA are further guided to the types of evidence they could obtain, such as network traffic captures, device logs, server logs, system logs and malicious binaries. Moreover, since the various data sources of SCI were already determined and implemented by the chosen vendors, retrieving the evidence would be significantly less challenging for SCI and LEA as methods or capabilities to retrieve these types of evidence should have been supported or documented as these SCI are constructed. In contrast to the instance discussed in the preceding paragraph, DFI and LEA enjoy the benefit of being prepared (knowing potential threats, types of cybercrime related to SCI, and types of evidence that could be collected). They can commence investigations effectively, efficiently, and with confidence.

\begin{figure}[ht]
\includegraphics[width=\columnwidth]{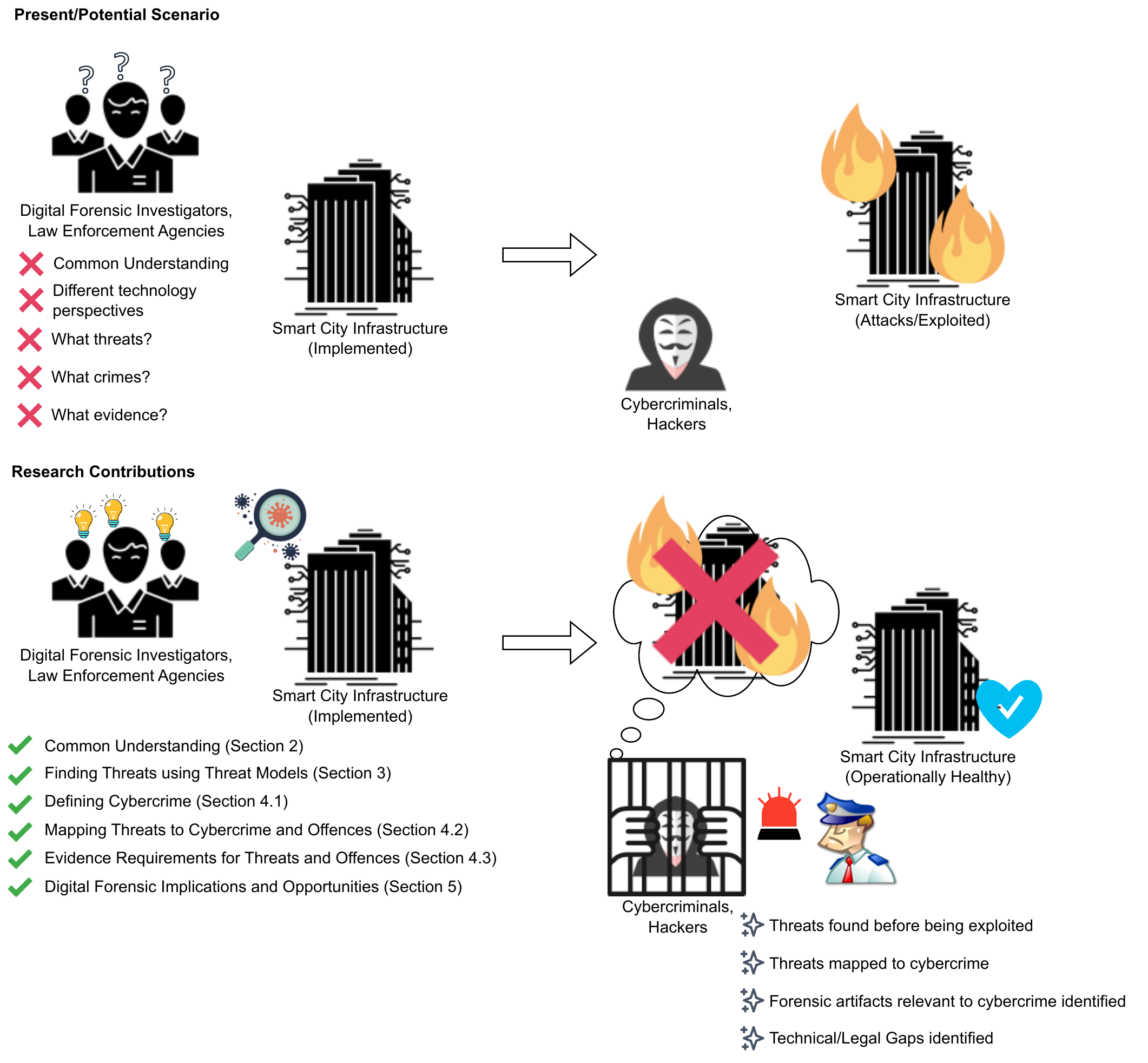}
\centering
\vspace{-15pt}\caption{Overview of our contribution, specific sections are highlighted}\vspace{-18pt}
\label{fig:5_DF_Implication}
\end{figure}

~\autoref{fig:5_DF_Implication} depicts the potential scenario faced by DFI and LEA and also shows how our proposed work can resolve their future challenges preemptively. 

\subsection{Research Opportunities for Digital Forensics} \label{Research_Opportunities}

Based on our research performed in Section~\ref{Background} to Section~\ref{SmartCityCyberCrime}, we identified a few key technical research opportunities for SCI digital forensics. These are 
outlined in the following: 

\subsubsection{Identification and Obtaining Dark Data} \label{DarkData}

Drawing a parallel connection with dark matter from Physics, Hand contends that missing data is dark data~\cite{HAND_2020}. Hand further postulates that there are 15 types of dark data, and such data is vital, although they have not been noted since the dark data could have a huge impact on corresponding analysis and actions~\cite{HAND_2020}. The current state of digital forensics on SCI exactly presents such a scenario. There is nothing wrong with governments wanting to implement SCI to improve the lives of citizens, but the corresponding identification of usable digital artifacts when cybercrime occurs has not been fully considered. Our research contributions (defining SCI, open-sourcing our Smart City Threat Model and identification of SCI evidence in Sections~\ref{Background} to~\ref{SmartCityCyberCrime}) attempts to address Dark Data-Type (DD-Type) 1 and 2 highlighted by Hand~\cite{HAND_2020}. DD-Type 1 refers to data we know are missing, while DD-Type 2 refers to data we do not know are missing~\cite{HAND_2020}.

The premise of our research was mainly driven by DD-Type 2 as we strove to identify the data we did not know that was missing in SCI digital forensics. After determining that there was a lack of a standard SCI definition, we formalize the SCI to facilitate understanding the threats, SCI cybercrime and subsequently, identifying SCI evidence sources. Consequently, certain DD-Type 2 data related to SCI became DD-Type 1, which were presented in our research. Given the nascent nature of SCI digital forensics, it is likely that other forms of dark data suggested by Hand exist in SCI deployments ~\cite{HAND_2020}. Additionally, since various governments may customize SCI depending on their use cases, the plethora of dark data could differ, which offers a wealth of technical research opportunities. Having said that, our proposed SCI threat model, identified threats, cybercrime and possible evidence provide a baseline for researchers, DFI and governments to refer to.

\subsubsection{Enhancing Evidence Capture in SCI} \label{SCI_Evidence_Automation_Capture}

Although we have suggested possible evidence sources and types in~\autoref{table:6}, evidence retrieval related to cybercrime will require human intervention. It will be a significant challenge to capture and identify the needed evidence at the right moment, especially when it is hard to predict when and where cybercrime will occur. A naive way would be to capture and store all evidence sources indefinitely, retrieving them when such evidence is required to investigate cybercrime. Unfortunately, such actions would not be economically feasible for governments, given the associated economical cost required to store the evidence. 

A possible solution to this dilemma could be using automated system processes to store evidence related to an ongoing cybercrime. However, this would require the engineering cooperation of the respective SCI vendors since this feature requirement appears to be an optional choice and not related to SCI core functionality. In the absence of vendor support, academic researchers could also possibly create tools or research on techniques that could be used in SCI to facilitate the investigation.

Another possible solution could be the usage of Machine Learning (ML) and Deep Learning (DL) techniques to process the various evidence sources and types to identify potential cybercrime that could have occurred. Such approaches have been explored in general cybersecurity and intrusion detection research~\cite{ML_DL_Cyber_Xin_2018, Dilara_2021}. With the support of ML/DL techniques, the strain of sifting through tremendous amounts of potential evidence could be reduced greatly. Investigation support could also be provided, though at this point of time, human supervision would still be a necessary thing due to the probabilistic nature of ML/DL tools. 
\subsubsection{SCI Evidence Storage and Management} \label{SCI_Evidence_Storage_and_Management}

It is not unusual for DFI to work with large amounts of data. The challenge of managing and working with large volumes of data and evidence has been highlighted in traditional digital forensic investigations~\cite{Quick2018}. In the case of SCI, the potential amount of varied data gathered for forensic investigation could be  tremendous, given the complexity and scale of SCI (with reference to~\autoref{table:6}). Researchers could adopt technical and management perspectives to tackle this challenge.

From a technical perspective, there are a few potential further research directions for forensic evidence storage. Firstly, some efforts could be directed towards a new medium of storage media or format where much more data could be stored resiliently and reliably. Alternatively, a suitable data compression mechanism could be explored to optimize and reduce evidence storage. Finally, the amount of evidence gathered within SCI could be reduced by applying data deduplication techniques. There has been open-sourced data deduplication frameworks such as DeFrame~\cite{Feng2022} and data deduplication extensions to open-sourced forensic tools such as The Sleuth Kit (TSK) proposed~\cite{Izabela_2022}. 

From a management perspective, researchers could examine various data policies related to SCI forensic evidence, such as storage, usage, access and acquisition vis-à-vis cybercrime investigation requirements. For example, it could be dictated under SCI evidence processing policies that data deduplication has to be carried out to reduce the potential amount of data that has to be examined and stored. Further policies such as access and storage would have to be determined ideally in partnership with LEA and governmental agencies.

\subsection{Legal Opportunities for Digital Forensics} \label{Legal_Opportunities}

On top of technical opportunities, we also identified some legal opportunities that could facilitate SCI cybercrime investigation based on our work in Section~\ref{Background} to Section-\ref{SmartCityCyberCrime}.

\subsubsection{Legal Jurisdiction of Digital Forensic Investigation} \label{DFI_Legal_Jurisdiction}

The availability of technology and access to the internet facilitates cybercrime. As such, various SCI systems could potentially be accessed or probed by international users. Adversaries and cybercriminal groups could capitalize on such opportunities to commit crimes and sabotage SCI. Even if SCI systems had their access restricted to a user's geolocation, adversaries located elsewhere could breach into a local victim's computer to access SCI. 

Since cybercriminals could be located in an international location, it may be challenging for local legal jurisdiction to prosecute  cybercriminals located in another country in the absence of reciprocal legal arrangements. However, LEA should consider the possibility of exercising their rights to extraterritorial jurisdiction. Gillespie suggests that there are four possible grounds in which LEA could apply extraterritorial jurisdiction~\cite{GillespieAlisdair2019C:KI}:

\begin{enumerate}
\item \textbf{Universal jurisdiction:} Reserved for the most grievous offences that was committed. The concept of this jurisdiction was that any nation could charge perpetrators regardless of their nationality and location where the crime was committed~\cite{aust_2010}. For example, perpetrators who interfere with the usual operations of SCI which leads to massive loss of human lives could be potentially considered under this category.  

\item \textbf{Protective principle:} The protective principle proposes to allow countries seek extraterritorial jurisdiction to defend the victim country's state interests and security~\cite{aust_2010}. For example, perpetrators who attempt cyberattacks on governmental or military assets owned by international governments could fall under this category.

\item \textbf{Active personality principle:} This form of extraterritorial jurisdiction is based on the the perpetrator's nationality and/or place of residence, and legal action could commence from there.

\item \textbf{Passive personality principle:} Similar to the active personality principle, this extraterritorial jurisdiction differs in that legal action commences instead from the victim's nationality and/or place of residence. 
\end{enumerate}

\subsubsection{International Cooperation in Digital Forensics} \label{DFI_International_Cooperation}

We discussed the possibility of extraterritorial jurisdiction in Section~\ref{DFI_Legal_Jurisdiction}. However, it would not be possible without a global consensus. The International Criminal Court (ICC) could be a potential platform where serious cybercrimes are judged, but it was observed that the ICC usually handles crimes against humanity~\cite{GillespieAlisdair2019C:KI}. 

A potential avenue could be the Budapest Convention on Cybercrime where the Second Additional Protocol to the Convention was recently put up for signing with the various parties and signatories~\cite{Budapest_Convention_2nd_Proto_2022}. Although the current scope of cooperation was limited to domain name and subscriber information, there was also consideration for extraterritorial cooperation via joint investigations and disclosure of electronic evidence~\cite{Budapest_Convention_2nd_Proto_2022}. However, the effectiveness of the Budapest Convention is only as effective as the parties and signatories who sign the Convention.
\section{Limitations} \label{Limitations}

Our proposed SCI threat model and the corresponding SCI threats, cybercrime and potential evidence are dependent on Definition~\ref{SCI_Definition}, the STRIDE threat model and the four ISO standards mentioned in Section~\ref{SmartCityDefinition}. There is a possibility that governments vested in alternative types of SCI (such as proprietary SCI not based on Definition~\ref{SCI_Definition}) would find a lack of alignment in our work. Nonetheless, the foundation of Definition~\ref{SCI_Definition} is based on the goals of an intergovernmental organization (17 United Nations SDGs) and several standards from an international standards body (ISO). Therefore, Definition~\ref{SCI_Definition} is technology agnostic and broad. While government agencies are free to select an SCI that best fits their needs, we believe that Definition~\ref{SCI_Definition} can assist governments in establishing a solid foundation in cybercrime prevention.

We used Microsoft TMT to create our threat model. As such, there is a dependency on a third-party tool. However, we believe the advantages outweigh the disadvantages of relying on a third-party tool. Threat modeling could be a herculean effort for complex systems such as SCI. Compared to manually creating threat models or using visualization tools with limited automation while creating threat models, the Microsoft TMT could track the list of identified threats, generate a summary report of threats and even provide suggested ways to mitigate identified threats. 

The possible evidence sources and types in~\autoref{table:6} are dependent on the threat model. In our research, we have limited the depth of our threat model at Level 1 as it was intended to serve as a baseline template for interested adopters. If we went down to more granular levels (e.g. Level 2 or 3) as specified by Microsoft, we would risk being overly specific and may outline redundant threats and evidence. Hence, although~\autoref{table:6} comprehensively maps threats on SCI to possible evidence sources and types, our threat model does not entirely outline all potential threats faced by SCI. Nonetheless, our baseline threat model provides a platform for extending towards different SCI customizations made by a 
specific organization or vendor.
\section{Related Work} \label{RelatedWork}

In our preparatory work for the paper, we performed a literature review of state-of-the-art works with the keywords ``Smart City'', ``Smart Cities'', ``Smart City Digital Forensics'', ``Cybercrime'', ``Threat Modeling'' and ``Threat Models''. The keywords were used in our searches via Google Scholar, journal databases such as ScienceDirect, IEEE and Association of Computing Machinery (ACM), and search engines such as Google. We manually checked the literature for the relevancy of the topics discussed in this paper and excluded papers unrelated to our research goals.

The idea of cybercrime had been widely discussed on several platforms but was first formally defined and classified by Gordon and Ford~\cite{GordonSarah2006Otda}. Gordon and Ford described cybercrime as any crime that was aided or executed via electronic devices such as computers or devices~\cite{GordonSarah2006Otda}. They also further classified cybercrime into Type I and II crimes - the former is often supported by malicious software such as keyloggers, viruses and vulnerabilities, while the latter uses day-to-day software and protocols. Type I cybercrime also usually happens as a specific event from the victim's perspective, while Type II cybercrime could repeatedly happen from the user's point of view. Gordon and Ford further argues that pure Type I or II cases are rarely observed and that cybercrime is usually broad\cite{GordonSarah2006Otda}. Our work, particularly~\autoref{table:5}, aims to resolve this conundrum by first identifying and mapping the possible threats to related offences. This prevents any confusion faced by DFI and LEA, albeit targeted specifically towards SCI-related cybercrime.

There have also been calls for digital forensics to evolve. MacDermott et al.~\cite{MacDermott_2018} highlighted that new ways to investigate IoT-related crime have become a growing necessity - threats and criminal abuses in smart cities are increasing while securing the infrastructure was a challenge. Moreover, MacDermott et al.~\cite{MacDermott_2018} further stated that new threats would have to mitigate by a new generation of digital forensics and best practices to ascertain and identify evidence in an evolving regulatory landscape concurrently. Our research precisely attempts to address this problem by combining digital forensics and industry best practices (e.g. threat modeling systems before implementation) to identify threats preemptively and to highlight cybercrime and needed digital evidence for investigation.

Baig et al.~\cite{BaigZubair2017} proposed a study of digital forensics and cybersecurity in smart cities based on NIST's smart city model. However, their study was only based on three categories (smart environments, smart living, smart mobility) out of the six categories mentioned in the NIST smart city model (smart economy, smart governance and smart people were not discussed). As such, this may lead to the misconception by researchers, DFI and LEA that smart cities only comprise smart environments, smart living and smart mobility. Meanwhile, our paper fully follows all clauses and elements of data indicators listed in the ISO standards we referenced. In addition, our paper attempts to define smart cities at an even lower layer of abstraction while embracing conformity to international standards (\textit{i.e.,} Definition~\ref{SCI_Definition}). Finally, governments will likely implement SCI infrastructure in phases. Our work can factor in technology evolution and facilitates requirement changes by governments (if necessary), yet provides DFI and LEA with the ability to prepare for cybercrime on existing SCI infrastructure.

A digital forensic model to identify interconnectivity between things was proposed by Kim et al.~\cite{KIM2023301499}, where the authors proposed an improved framework to identify evidence in interconnected devices. While the proposed model is informative, it lacks the context of SCI and assumes that DFI and LEA have an in-depth knowledge of SCI. It also requires users to follow the process iteratively. On the other hand, our threat model lists all possible threats within SCI that are grouped by data indicators via threat modeling immediately. The possible offences, forensic evidence types and sources based on the threats are then linked together and presented in our research. By using our threat modeling template, governments, DFI, and LEA could obtain personalized guidance and direction in identifying relevant threats and sources of forensic data. If certain data indicators were not required, they could be trimmed away from the template.

Kavallieratos et al. investigated threats in connected smart homes using STRIDE and Microsoft TMT~\cite{Kavallieratos2019}.  
They acknowledged the dynamic nature of smart, interconnected devices but only outlined six smart home scenarios and solely determined threats. Moreover, Kavallieratos et al. did not adhere to Microsoft's best practices for threat modeling and failed to model their threat models according to appropriate context levels~\cite{Microsoft_TM_DFD_Layer_2022}. As such, it is possible that Kavallieratos et al. did not identify all existing threats in their scenarios.  

Countries have attempted to specify technical references and standards for smart cities/nations to enrich their citizens' lives. For example, in Technical Reference (TR) 47, Singapore attempted to specify a technology-agnostic reference architecture for IoT and sensor networks~\cite{TR47}. The motivation of the TR was to create an anticipatory government that could offer integrated city services while allowing citizens to enhance their lives with data~\cite{TR47}. Other than technical requirements specified by Singapore, it was observed that TR 47 also mentioned the need for incident response and investigations in the architecture~\cite{TR47}. Our work could provide the investigative context and guidance required from a data indicator perspective.

In summary and to the best of our knowledge, we provide the first systematic SCI threat model that is technology-agnostic and adopts data indicators that have been internationally agreed upon (i.e., ISO standards). Instead of focusing on the technological aspect of SCI, we based our SCI on indicators that reflected the quality of life and used the corresponding data indicators in our SCI.
\section{Discussion and Conclusion} \label{Conclusion}

Determining SCI threats, cybercrime, and retrieving related forensic evidence is not a trivial task for DFI and LEA. Even at a more fundamental level, the definition of SCI could differ due to different needs and perspectives. Our research work led us to the following key observations:

\begin{enumerate}
    \item \textbf{SCI Definition.} A common definition of SCI supported by international standards helped to reduce ambiguity when we had to prepare for our threat modeling endeavors. We could be technology-agnostic in our threat model by viewing SCI as an outcome-based and data-driven construct compared to a purely technological and role-based construct. With a standard definition and understanding of SCI, DFI and LEA would be better prepared to handle SCI investigations.
    \item \textbf{SCI Threats, Cybercrime and Related Evidence.} As Horsman has succinctly pointed out, DFI and LEA have already had their hands full with ongoing cybercrime cases and are hard-pressed to perform further research in emerging technologies~~\cite{HORSMAN2019100003}. We believe that adversaries and cybercriminals are very likely to target SCI due to their novelty and the data that flows through these systems. Before SCI is fully implemented and deployed, we preemptively highlight vital details that DFI and LEA are likely to use in their course of investigation and workflows. Based on a standard definition of SCI, we derived potential threats and mapped them to possible offences. We then further mapped the threats to possible evidence types and sources. With these details, DFI and LEA will have some guidance when they are required to conduct investigations related to SCI.
	\item \textbf{SCI Technical and Legal Opportunities for Digital Forensics.} Although our research attempts to get DFI and LEA up to speed about SCI, there are still multiple opportunities for the digital forensics community to explore. We hope that the digital forensics community can work together in looking into the opportunities identified - a two-pronged approach from both a technical and legal perspective would better assist DFI and LEA, who would have to contend with SCI investigations in the future.
\end{enumerate}

We have provided a threat model template freely available for usage by interested parties to model their SCI and discover potential threats. Following that, we also listed possible digital evidence that DFI and LEA could use to investigate cybercrime based on these threats. We believe that our work in this paper will serve as a critical foundation in aiding DFI and LEA in investigating SCI cybercrime, while also factoring in the goal of improving citizens' lives by aligning our threat model with the UN sustainability issues and goals as specified in ISO37101:2016.

\smallskip\noindent
\textbf{Availability:} Our threat models are publicly available at \url{https://github.com/poppopretn/SmartCityThreatModel} for investigation and further research. 
\section{Acknowledgements}

This work is partially supported by the NRF National Satellite of Excellence in Trustworthy Software Systems (Project no. NSOE-TSS2021-01 and NSOETSS2020-03).

{
	\bibliographystyle{unsrt}	
	\bibliography{references}
}


\end{document}